\documentclass[preprint2]{aastex}

\shorttitle{Methods and performances of VIMOS IFU data reduction}
\shortauthors{Zanichelli et al.}

\begin{document}

%% LaTeX will automatically break titles if they run longer than
%% one line. However, you may use \\ to force a line break if
%% you desire.

\title{The VIMOS Integral Field Unit: data reduction methods 
	and quality assessment}

%% Use \author, \affil, and the \and command to format
%% author and affiliation information.
%% Note that \email has replaced the old \authoremail command
%% from AASTeX v4.0. You can use \email to mark an email address
%% anywhere in the paper, not just in the front matter.
%% As in the title, use \\ to force line breaks.

\author{A. Zanichelli\altaffilmark{1},
B. Garilli\altaffilmark{2}, M. Scodeggio\altaffilmark{2},
P. Franzetti\altaffilmark{2}, D. Rizzo\altaffilmark{3},
D. Maccagni\altaffilmark{2}, R. Merighi\altaffilmark{4},
J. P. Picat\altaffilmark{3}, O. Le F\`evre\altaffilmark{5},
S. Foucaud\altaffilmark{2},
D. Bottini\altaffilmark{2}, V. Le Brun\altaffilmark{5},
R. Scaramella\altaffilmark{1}, L. Tresse\altaffilmark{5},
G. Vettolani\altaffilmark{1}, C. Adami\altaffilmark{5},
M. Arnaboldi\altaffilmark{6}, S. Arnouts\altaffilmark{5},
S. Bardelli\altaffilmark{4}, M. Bolzonella\altaffilmark{7},
A. Cappi\altaffilmark{4}, S. Charlot\altaffilmark{8,9},
P. Ciliegi\altaffilmark{4},
T. Contini\altaffilmark{3},
I. Gavignaud\altaffilmark{3,10}, L. Guzzo\altaffilmark{11},
O. Ilbert\altaffilmark{5}, A. Iovino\altaffilmark{11},
H. J. McCracken\altaffilmark{9,12}, B. Marano\altaffilmark{7},
C. Marinoni\altaffilmark{5}, G. Mathez\altaffilmark{3},
A. Mazure\altaffilmark{5}, B. Meneux\altaffilmark{5},
S. Paltani\altaffilmark{5}, R. Pell\`o\altaffilmark{3},
A. Pollo\altaffilmark{11}, L. Pozzetti\altaffilmark{4},
M. Radovich\altaffilmark{6}, G. Zamorani\altaffilmark{4},
E. Zucca\altaffilmark{4}
}

%% Notice that each of these authors has alternate affiliations, which
%% are identified by the \altaffilmark after each name.  Specify alternate
%% affiliation information with \altaffiltext, with one command per each
%% affiliation.
\altaffiltext{1}{IRA-INAF - via Gobetti, 101, I-40129 
Bologna, Italy; {\it a.zanichelli@ira.cnr.it}}
\altaffiltext{2}{IASF-INAF - via Bassini, 15, I-20133 Milano, Italy}
\altaffiltext{3}{Laboratoire d'Astrophysique de l'Observatoire 
Midi-Pyr\'en\'ees (UMR 5572) - 14, avenue E. Belin, F31400 Toulouse, France}
\altaffiltext{4}{INAF-Osservatorio Astronomico di Bologna - via Ranzani, 1, 
I-40127 Bologna, Italy}
\altaffiltext{5}{Laboratoire d'Astropysique de Marseille, UMR 6110 
CNRS-Universit\'e de Provence,  BP8, 13376 Marseille Cedex 12, France}
\altaffiltext{6}{INAF-Osservatorio Astronomico di Capodimonte - via Moiariello,
16, I-80131 Napoli, Italy}
\altaffiltext{7}{Universit\`a di Bologna, Dipartimento di Astronomia - via 
Ranzani, 1, I-40127 Bologna, Italy}
\altaffiltext{8}{Max Planck Institut fur Astrophysik, 85741, Garching, Germany}
\altaffiltext{9}{Institut d'Astrophysique de Paris, UMR 7095, 98 bis Bvd Arago,
75014 Paris, France}
\altaffiltext{10}{European Southern Observatory, Karl-Schwarzschild-Strasse 2,
D-85748 Garching bei Munchen, Germany}
\altaffiltext{11}{INAF-Osservatorio Astronomico di Brera - via Brera, 28, 
Milano, Italy}
\altaffiltext{12}{Observatoire de Paris, LERMA, 61 Avenue de l'Observatoire, 
75014 Paris, France}

%% Mark off your abstract in the ``abstract'' environment. In the manuscript
%% style, abstract will output a Received/Accepted line after the
%% title and affiliation information. No date will appear since the author
%% does not have this information. The dates will be filled in by the
%% editorial office after submission.

\begin{abstract}
With new generation spectrographs integral field spectroscopy is becoming a widely used observational technique.
The Integral Field Unit (IFU) of the VIsible Multi--Object Spectrograph (VIMOS)
on the ESO--VLT allows to sample a field as large as 
$54\arcsec \times 54\arcsec$ covered by 6400 fibers coupled with 
micro-lenses. 
We are presenting here the methods of the data processing software developed 
to extract the astrophysical signal of faint sources from the VIMOS IFU 
observations. We focus on the treatment of the fiber-to-fiber relative
transmission and the sky subtraction, and the dedicated tasks we have built to
address the peculiarities and unprecedented complexity of the dataset. 
We review the automated process we have developed under the VIPGI data 
organization and reduction environment \citep{Scod05}, along with the quality 
control performed to validate the process. The VIPGI-IFU data processing 
environment is available to the scientific community to process VIMOS-IFU data 
since November 2003.
\end{abstract}

%% Keywords should appear after the \end{abstract} command. The uncommented
%% example has been keyed in ApJ style. See the instructions to authors
%% for the journal to which you are submitting your paper to determine
%% what keyword punctuation is appropriate.


\section{Introduction}

Integral field spectroscopy (IFS) is one of the new frontiers of modern 
spectroscopy.
Large, contiguous sky areas are observed to produce as many spectra as there 
are spatial resolution elements sampling the field of view. Integral field 
units (IFU) use micro-lenses, eventually coupled to fibers, or all reflective 
image slicers \citep{cont00,prieto00} to transform the 2D field of view at the
telescope focal plane into a long slit, or a set of long slits, at the 
entrance of the spectrograph.
After the dispersing element, the contiguous spatial sampling and spectra 
produce a 3D cube containing ($\alpha$, $\delta$), and $\lambda$ information 
\citep[see e.g.][]{bacon95,alli98,bacon01,alli02}.

The application of integral field spectroscopy to astrophysical studies may 
overcome many of the limitations posed by classical long slit or multi-slit 
spectroscopy.
It is particularly powerful to study objects with a complex 2D 
distribution of the spectral quantity to be measured. It can give a clear 
advantage over classical long slit or multi-slit spectroscopy 
because in one single observation it samples the object to be studied, 
whatever its complex shape, and because it collects all the emitted light 
(no slit losses).
For instance, measuring the redshifts of galaxies in
the core of distant clusters of galaxies is much more efficient with
the IFS technique than with 
multi-slit spectrographs because of the closely packed geometry of the core 
galaxies. 
IFS techniques are also a very powerful tool in the study 
of the internal dynamical structure of galaxies.
Large-scale kinematical studies of galaxies have been strongly limited by the 
insufficient 
spatial sampling of long slit spectroscopy,
until the first studies with integral field spectrographs like TIGER 
\citep{bacon95} appeared, followed by large samples of galaxies observed with 
SAURON \citep{bacon01,Emsell04}.
IFUs are well suited also for observations of low surface brightness galaxies:
the slit loss problem faced by conventional spectrographs does not exist with 
IFUs, and such faint, extended objects are less difficult to detect. 

The role of IFS is widely recognized today as a key technology to help solve 
some of the most fundamental questions of astrophysics, but dealing with the 
data obtained with integral field spectrographs is still a challenging task.
On one side, the reduction of data taken with fiber-based integral field 
spectrographs presents some peculiar aspects with respect to classical slit 
spectroscopy: for instance, variations in fiber relative transmission must be 
properly treated and sky subtraction is a crucial step for many of these 
spectrographs.
On the other side, the huge amount of data obtained with even one night of 
observations with the new-generation integral field spectrographs, like the 
VIMOS Integral Field Unit \citep{Bonne03}, makes it impossible to process them
by hand.
A new approach is required, based on the implementation of dedicated reduction
techniques inside an almost completely automated pipeline for data processing.

The Integral Field Unit of VIMOS is the largest IFS ever built on an 8m class 
telescope. 
The high spectra multiplexing of VIMOS required the development of VIPGI,
a semi--automatic, interactive pipeline for data reduction \citep{Scod05}.
The core reduction programs that constitute the main data processing 
engine for the reduction of VIMOS data have been developed as part of the 
contract between the European Southern Observatory and the VIMOS Consortium. 
Such data reduction software is now part of the on-line automatic pipeline for
VIMOS data at ESO. 
With VIPGI we have kept the capability of these core reduction programs for
a fast reduction process, but we have also added interactive tools that make 
it a careful and complete science reduction pipeline.
VIPGI capabilities include dedicated plotting tools to check the quality and
accuracy of the critical steps of data reduction, a user--friendly graphical
interface and an efficient data organizer.
The VIPGI interactive pipeline has been offered to the scientific community 
since November 2003 by the VIRMOS Consortium, to support observers through 
the data reduction process.

In this paper we describe the peculiar aspects, the principles of operations
and the performances of the VIMOS IFU data reduction pipeline implemented 
within VIPGI.
In Sect. \ref{sec:IFU} we describe the VIMOS IFU and in Sect.\ref{sec:genred}
we give the motivations leading to the development of a new, dedicated 
pipeline for VIMOS IFU data while describing the general concepts of data 
analysis. 
Some aspects specific of the VIMOS IFU data processing and analysis are 
discussed in more detail in Sects. \ref{sec:trace} through \ref{sec:jitter}, 
together with examples of the results obtained.

\section{The VIMOS Integral Field Unit}\label{sec:IFU}

VIMOS (VIsible Multi Object Spectrograph) is a high multiplexing 
spectrograph with imaging capabilities installed on the third unit of the Very
Large Telescope and designed specifically to carry out survey work
(Le F\`evre et al. 2002).

A detailed description of the VIMOS optical layout and of the MOS observing 
mode in particular can be found in \citet{Scod05}.
VIMOS main features are the capability of simultaneously obtaining up to 800 
spectra in multi-slit mode and the availability of a microlens-fiber unit 
designed to perform integral field spectroscopy.
To achieve the largest sky coverage the VIMOS instrument has been split 
into four identical optical channels/quadrants, each acting as a classical
focal reducer. When in MOS mode, each channel samples $\approx 7\times8$ 
arcmin on the sky, with a pixel scale of 0.205 arcsec/pixel.
The MOS and Integral Field Unit modes share entirely the four VIMOS optical 
channels. However, the VIMOS optical path in IFU mode differs from the MOS 
one in three elements: the so-called IFU head, the fiber bundle and the 
IFU masks.

The IFU head is placed on one side of the VIMOS focal plane (see Fig. 1 in
Scodeggio et al. 2005) and consists of $6400$ microlenses organized in an 
$80\times 80$ array. Each microlens is coupled with an optical fiber.
Spatial sampling is continuous, with the dead space between fibers below
$10\%$ of the fiber-to-fiber distance.
The fiber bundle, which provides the optical link between the microlenses 
array of the IFU head and the VIMOS focal plane, is first split into 4 parts,
each feeding one channel, and then distributed over ``pseudo-slits'' carved 
and properly spaced into 4 masks. Output microlenses on the pseudo-slits 
restore the f/15 focal ratio needed as input to the spectrograph. 
The IFU masks are movable devices, when the IFU observing mode is selected 
they are inserted in the focal plane replacing the MOS masks.

The optical configuration of the IFU guarantees that field losses do not 
exceed $5\%$

\begin{figure*}
\resizebox{\hsize}{!}{\includegraphics{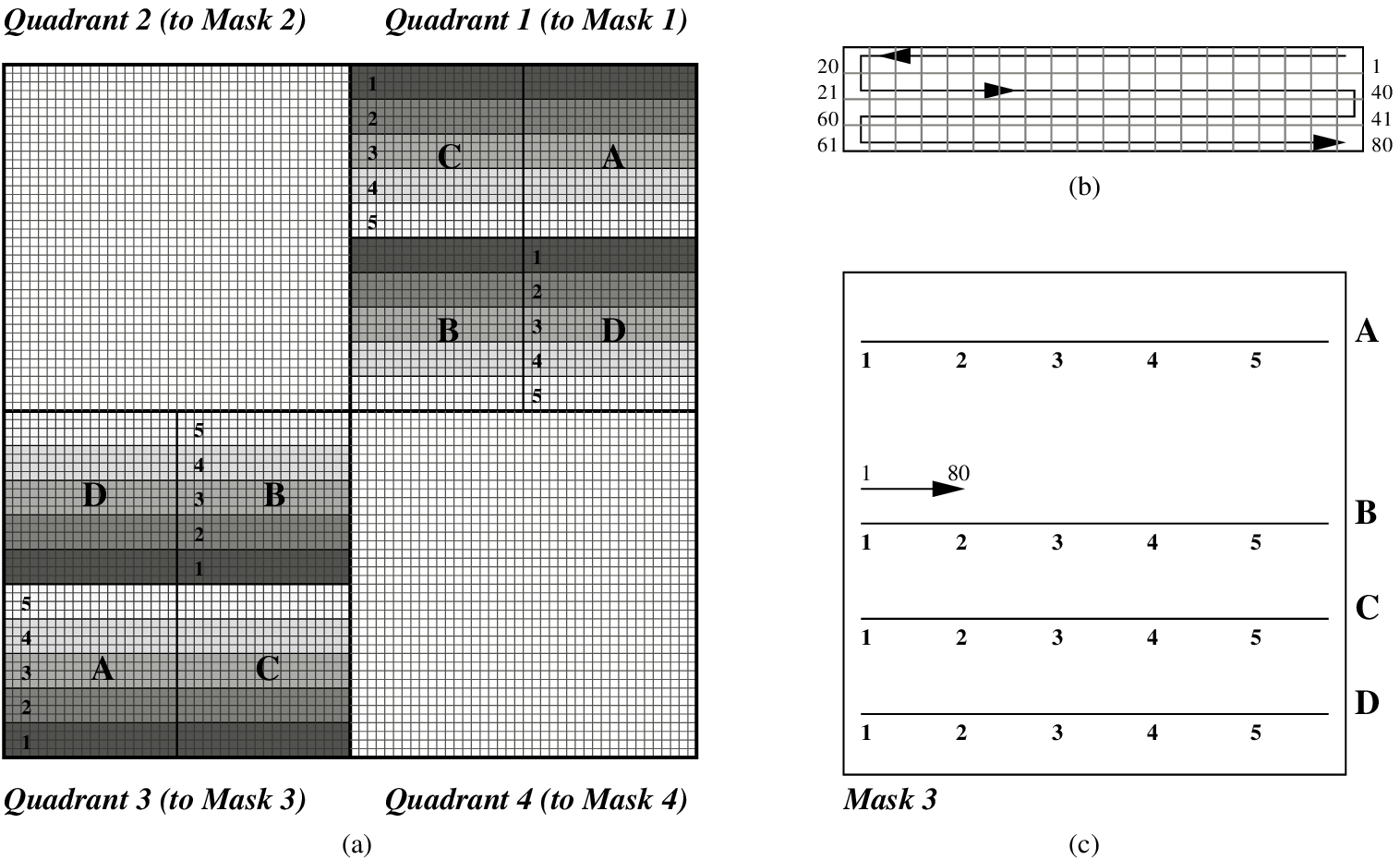}}
\caption{Geometrical layout of the VIMOS Integral Field Unit.
(a) The $80 \times 80$ microlenses array that form the IFU head. 
Each quadrant is associated with a $1600$ fibers bundle which conveys the 
collected light into one of the four VIMOS channels. 
For clarity, only details for quadrants 1 and 3 are shown.
Sub-bundles group fibers associated with contiguous microlenses on each
quadrant (the regions marked A, B, C, and D)
and feed the four pseudo-slits on the special IFU masks put
in the VIMOS focal plane. 
A fiber sub-bundle is in turn divided into 5 modules of 80 fibers each.
The fibers in each module are aligned onto the pseudo-slits according to a 
complex pattern, as can be seen in (b).
Panel (c) illustrates how the fiber modules are organized over the pseudo-slits
in the case of the IFU Mask no. 3.}
\label{fig1}
\end{figure*}

In  Fig. \ref{fig1} a schematic view of the IFU geometrical configuration
is given. The microlens array of the IFU head with superimposed the division
of the fiber bundle into the four VIMOS quadrants/IFU masks is described 
in Fig. \ref{fig1}~(a).
Fibers going to different pseudo-slits belonging to the same IFU mask 
are grouped in sub-bundles,  indicated with A, B, C, and D.
Each sub-bundle comprises five independent {\it modules} of $20 \times 4$ 
fibers: these are the ``fundamental units'' of the IFU bundle and are
marked with different gray levels.
The $20 \times 4$ fibers in a module are re-arranged  to form a linear array
of $80$ fibers on a pseudo-slit (see Fig. \ref{fig1}~(b) for an example).
Each pseudo-slit holds five fiber modules.
Fig. \ref{fig1}~(c) shows how the modules are organized over the pseudo-slits
in the case of the IFU mask corresponding to quadrant no. 3.

\begin{figure*}
\resizebox{\hsize}{!}{\includegraphics{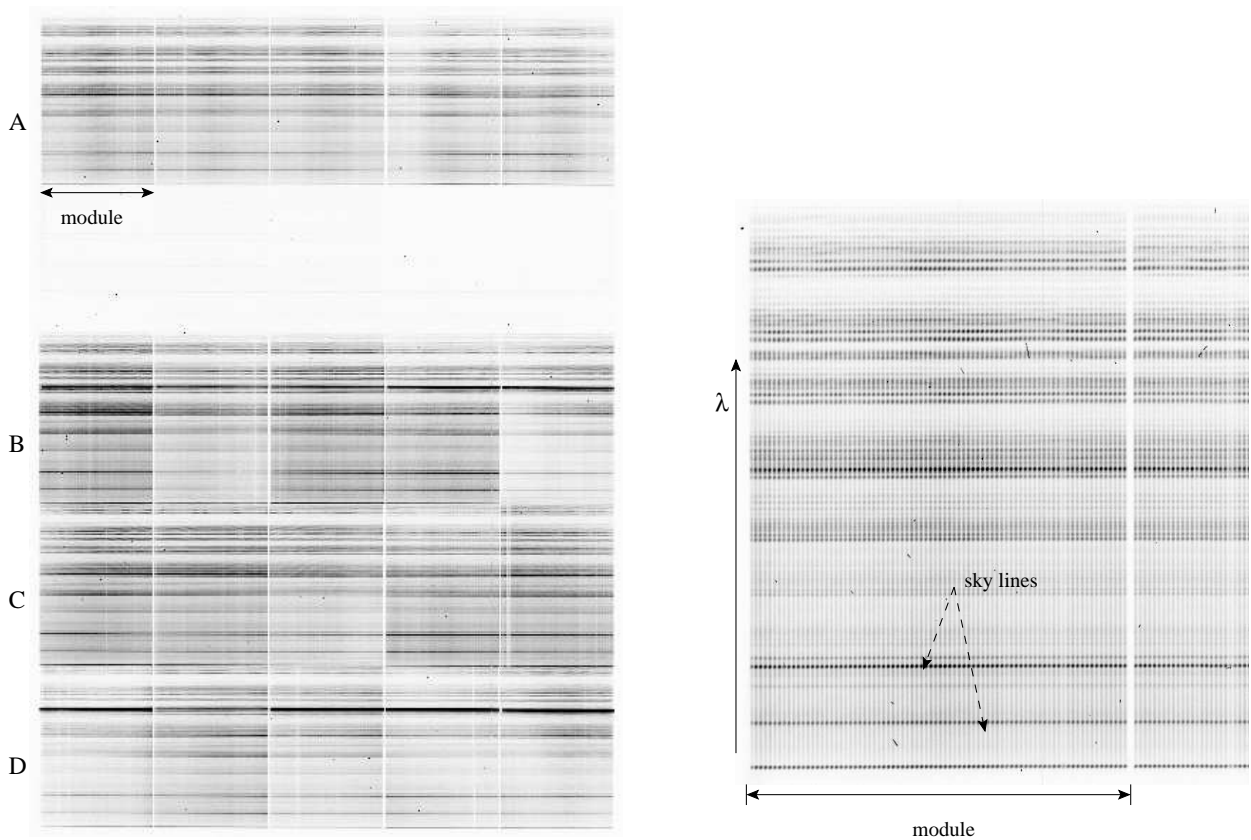}}
\caption{Left: example of an IFU exposure, only one of four 
quadrants is shown, the four pseudo-slits with $400$ spectra each are visible.
Each spectrum spans 5 pixels in the spatial direction. One fiber module 
belonging to pseudo-slit A is indicated in the left panel; a zoom on the 
$80$ spectra coming from this module is shown in the right panel.}
\label{fig2}
\end{figure*}

Contrary to what happens for MOS observations, the pseudo-slit positions on 
the IFU masks are fixed. This produces a fixed geometry of the spectra on 
the four VIMOS CCDs (see Fig. \ref{fig2} for an example).
The information on the correspondence between the position of a fiber in the 
IFU head and the position of its spectrum on the detector is one of 
the fundamental ingredients of the data reduction process and, together with 
other fiber characteristics, it is stored in the so-called ``IFU table'' 
(Sect. \ref{sec:genred}).

	\placefigure{fig1}

IFU observations can be done with any of the available VIMOS grisms (see 
Table 1 in Scodeggio et al. 2005). 
At low spectral resolution ($R\sim 200$), 4 pseudo-slits per quadrant 
provide $4 \times 400$ horizontally stacked spectra on each of the four 
VIMOS CCDs.
In the left panel of  Fig. \ref{fig2}~ the image of quadrant/CCD no. 3 in
an IFU exposure taken with the Low Resolution Red grism is shown, the 
four pseudo-slits holding $400$ spectra each are clearly visible. 
One of the fiber modules belonging to pseudo-slit A is indicated, and a zoom
on its $80$ spectra can be seen in the right panel.
At high resolution ($R\approx 2500$) spectra span a much larger number of 
pixels in the wavelength direction over the CCDs and only the central 
pseudo-slit on each mask can be used.
The complex rearrangement of fiber modules from the IFU head to the masks is
such that the four central pseudo-slits (marked as ``B'' in Fig. \ref{fig1})
map exactly the central part of the field-of-view.
This makes it possible to perform high spectral resolution observations  
while keeping the advantage of a contiguous field. A dedicated shutter is 
used to select only the central region of the IFU field-of-view.

Two different spatial resolutions, $0.67^{\prime\prime}$/fiber and 
$0.33^{\prime\prime}$/fiber, are possible thanks to a removable focal 
elongator that can be placed in front of the IFU head. The higher spatial 
resolution translates in a smaller field-of-view.
The sky area accessible to an IFU observation is thus function of the chosen
spectral and spatial samplings.
Table \ref{tab:IFU} summarizes the values for the IFU field size  
as a function of the allowed spectral and spatial samplings.

	\placefigure{fig2}

\section{The IFU Data Reduction}\label{sec:genred}

The IFU data reduction is part of the VIMOS Interactive Pipeline and
Graphical Interface (VIPGI).
For a detailed description of the VIPGI functionality as well as the 
handling and organization of VLT-VIMOS data we refer to \citet{Scod05}.

The realization of a dedicated pipeline for the VIMOS IFU is mainly 
motivated by two considerations:
1) the need to be independent from already existing software
environments, like IRAF or IDL ones, which was (at that time) 
a general ESO requirement for VLT instrument data  reduction pipelines;
and 2) some peculiarities of the instrument which required the 
development of new specific processing tools.
As an example, the continuous coverage of the field-of-view does not 
guarantee to have dedicated fibers for pure-sky observations during each 
exposure. A new tool for sky subtraction (Sect. \ref{sec:skysub})
has been developed, which does not require special observing techniques, 
like the ``chopping'' one, and thus does not impact on the observing 
overheads.

As for MOS, the starting point for the reduction of IFU data is the knowledge
of an instrument model (see Sect. 4 in \citet{Scod05} for details), 
i.e. an optical distortion model, a curvature model and a wavelength 
dispersion solution.
These models are periodically derived by the ESO VIMOS instrument 
scientists using calibration plan observations, and are stored in the image 
FITS headers as ``first guess'' polynomial coefficients.
First guess parameters are used as a starting point to refine the instrument 
model on  scientific data. With such an approach, the best possible 
calibration is obtained for each individual VIMOS exposure.
The refinement of first guess models is a fundamental step in IFU data
reduction, since the instrumental mechanical flexures are often 
a critical factor, see Sect. \ref{sec:locate}.

From the hardware point of view, each VIMOS quadrant is indeed a 
completely independent spectrograph, characterized by its own instrument 
model. For this reason IFU  data processing 
is performed on single frames, 
i.e. images from each quadrant are reduced separately, up to the creation
of a set of fully calibrated, 1D reduced spectra. 
The final steps of data reduction, like the creation of a 2D reconstructed 
image or the combination of exposures in a jitter sequence (Sect. 
\ref{sec:jitter}), are on the contrary performed only once all images from 
all the quadrants have been reduced.

A high degree of automation is achieved by means of auxiliary tables 
needed by the data reduction procedures. In the case of VIMOS IFU,
fundamental information is listed in the IFU Table. Starting from the 
instrument layout, this table gives the one to one correspondence between 
fiber position on the IFU head and spectra on the CCD, as well as other fiber
parameters like the relative transmission and the coefficients describing the
fiber spatial profile (see discussion in Sect. \ref{sec:skysub}).

\begin{figure}
\plotone{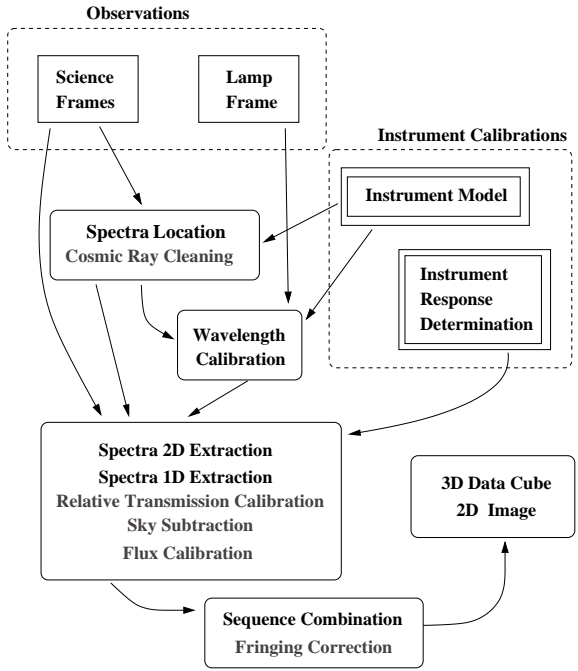}
\caption{Block diagram showing the various steps of the reduction of VIMOS 
IFU data. Optional steps are marked in grey.}
\label{fig3}
\end{figure}

The first steps of IFU data reduction are: tracing of spectra on the 
CCD, cosmic ray cleaning and wavelength calibration -- operations that lead to
the extraction of 2D spectra. 
Wavelength calibration is performed as in the MOS case, with an accuracy of 
the computed dispersion solution comparable to what obtained for MOS spectra 
\citep[see][for details]{Scod05}.
Extraction of 1D spectra is generally done with the usual \citet{Horn86}
extraction method, by means of spatial profiles determined for each fiber from 
the data themselves. 1D extracted spectra may be calibrated to correct 
for differences in fiber transmission and properly combined to determine and 
subtract the sky spectrum. Flux calibration may be applied as a final step.
If observations have been carried out using the shift-and-stare
technique, 1D reduced single spectra belonging to the same sequence 
may be corrected for fringing and properly combined in a data cube. Finally, a
2D reconstructed image is built.
A block diagram of the operations performed by the VIMOS IFU data reduction 
pipeline is shown in Fig. \ref{fig3}. In grey are marked those steps that 
have been left as an option. For instance, removal of cosmic ray hits as well
as relative transmission correction and sky subtraction are not strictly 
needed in the reduction of short exposures of spectrophotometric standard 
stars. Moreover, sky subtraction may not satisfactorily work in the case of 
very crowded fields (see Sect. \ref{sec:skysub}) and can be skipped.
Fringing correction is not necessary when the VIMOS blue grism is used, 
because its wavelength range is free from such an effect.

Key steps in the processing of VIMOS IFU data are the spectra location on the
CCDs, cosmic ray cleaning, cross-talk contamination and relative transmission
corrections, and sky subtraction.
In the following Sections we will focus our attention on these steps to 
clarify their impact on the data reduction and to motivate the need of 
dedicated reduction procedures, as well as to describe the adopted methods 
and to show the obtained quality.

	\placefigure{fig3}

\section{Extraction of VIMOS IFU spectra}\label{sec:trace}

The extraction procedure consists in tracing spectra on the CCDs, applying
wavelength calibration and extracting 2D and then 1D spectra.
The most critical aspects of spectra extraction are the accurate location of 
spectra positions on the detectors and the cosmic ray cleaning; the 
crosstalk effect is discussed in Sect. \ref{sec:ctcorr}.

\subsection{Locating spectra}\label{sec:locate}

When dealing with VIMOS IFU data, spectra location is a much more critical
step than in multi-slit mode. This is a consequence of two instrumental 
characteristics of the VIMOS spectrograph: mechanical flexures and spectra 
distribution on the detectors.

As already discussed in Section 2 of \citet{Scod05}, the overall instrumental
flexures induce image motion of the order of $\sim \pm 2$ pixels.
In IFU mode, additional and predominant sources of mechanical instability 
are the deployable IFU masks, and flexures are larger than in MOS mode.
Shifts of the order of $\pm 5$ pixels in the spectra positions between 
exposures taken at different rotation angles are typical, but values as large
as $11$ pixels have been observed.
Such values are comparable or larger than the spatial extent of a fiber 
spectrum on the VIMOS detectors (about $5$ pixels) and are strongly dependent
on the instrument rotator angle during the observations.

If shifts are comparable to the spatial size of a spectrum, it may become 
impossible to correctly determine the correspondence between a fiber on an 
IFU mask and its spectrum on the detector.
As can be seen in Fig. \ref{fig2}, the $\sim 10$ signal free pixels between 
modules are easily identifiable regions, and a simple user 
interface allows to set their (rough) position. 
Once an inter-module position is known, it is used as starting point from 
which, moving leftwards and rightwards, to measure the positions of the 80 
spectra belonging to adjacent modules. 
The spectra positions are traced with a typical uncertainty of $0.5$ pixels.
For each spectrum, a second degree polynomium is fit to these positions and 
its coefficients are stored in the so-called extraction table, to be used for
2D extraction.
With respect to the instrument model coefficients, the extraction table gives
a much more accurate description of the spectra location since it is 
``tuned'' on the data themselves.

The spectra location method for VIMOS IFU data has been extensively tested
on data taken with the different grisms and proved to be extremely robust in
correctly identifying fibers and tracing spectra, irrespective of the amount
of shifts induced by flexures and/or distortions.

\subsection{Cosmic Ray Cleaning}\label{sec:cosmclean}

Once spectra have been traced and  before the 2D spectra are extracted, 
cleaning from cosmic ray hits is performed.
This reduction step is of great importance to get a good relative transmission
calibration and sky spectrum determination, since the results of these tasks 
can be seriously affected by the presence of uncleaned cosmic rays altering 
spectral intensities.

In principle, the best way to remove cosmic ray hits would be to compare 
different exposures of the same field.
However, two reasons forced us to develop an alternative method: first, in
the case of VIMOS IFU image displacements due to mechanical flexures prevent
the direct comparison of pixel intensities to remove cosmic ray hits.
Secondly, we were interested in having a method general enough to be 
applicable also in the case of single exposures of bright objects.
Taking into account these considerations we developed an algorithm which 
works on single frames and whose principles of operations are applicable also
to data taken with spectrographs other than VIMOS.

\begin{figure}
\plotone{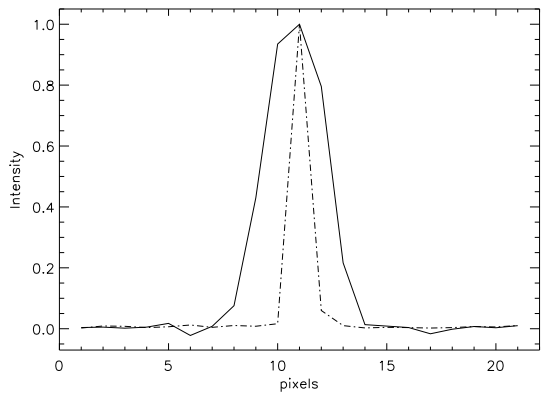}
\caption{Sky line profile along the dispersion direction (continuum 
line), superimposed to the profile of a cosmic ray (dot-dashed line).}
\label{fig4}
\end{figure}

Compared to other existing tools for single-frame cosmic ray removal (e.g. 
the {\it cosmicrays} routine in IRAF), the method implemented in the VIMOS 
IFU pipeline is different because it relies on the hypothesis that - along 
the wavelength direction - spectra show a smooth behaviour in their 
intensities. In presence of emission lines or sky lines their intensity 
{\it gradients} will be smooth enough to be distinguishable, when compared 
with the abrupt changes generated by cosmic rays. This is shown in Fig. 
\ref{fig4}, where the profile of a sky line as obtained with the Low 
Resolution Blue grism is superimposed to the profile of a cosmic ray. Both 
profiles are cut along the wavelength direction.

	\placefigure{fig4}

Each IFU spectrum is traced along the CCD column by column and the
intensity gradient array is computed and analyzed to search for sharp 
discontinuities.
Intensity gradient arrays are inspected by sliding along them with a small
"window" (of the order of 20 pixels in length) inside which the local mean 
and rms values are computed and used to discard discrepant gradient values, 
likely to be due to the presence of a cosmic ray.
The median and rms are re-computed on the clipped gradient window data and 
used to build the actual upper and lower thresholds for cosmic ray detection
in the gradient window. Window size and number of sigmas to compute 
thresholds are user selected parameters.

Multiple pixel hits are hard to distinguish from true emission lines.
In the attempt to clean as many cosmic rays as possible, without altering 
emission line shapes, before replacing the ``suspect pixel value'' we perform
a further check. 
The mean intensity in the local window is compared with the single pixel
intensities of the corresponding columns of the adjacent fibers.
Because of the window dimensions (typically 20 pixels) over which the mean 
intensity is computed, such values should not be significantly different, 
except when an emission line is present. 
When a significant difference is found, this is interpreted as presence of an
emission line, and an average of the intensities in the two comparison 
spectra  is used as replacement value.
On the contrary, if there is no difference, the intensity of pixels
identified as cosmic rays is replaced with the local mean value. 

Tests on spectra taken with the different VIMOS grisms showed that among all 
the pixels classified as cosmic ray hits on the first pass, only about 
$0.3\%$ actually belong to spectral lines. 
In all these cases, the comparison with neighbouring spectra has NOT
confirmed the ``cosmic ray hypothesis'', and data have not been incorrectly 
replaced.

It could be argued that emission lines from pointlike sources, occupying one
fiber only, would be deleted by this method. In reality, given the median 
seeing in Paranal (0.8 $\arcsec$) and the fiber dimension on sky (0.67 
$\arcsec$), such an eventuality is extremely rare.

The cosmic ray hits removal method implemented in the VIMOS IFU pipeline
has been extensively tested and proved to be very efficient in cleaning both 
high and low spectral resolution data. On average, $\sim 90\%$ of the cosmic 
rays are removed: for cosmic ray hits spanning 1-2 pixels, the removal 
success rate is $99\%$, while more complex, extended hits are cleaned with a
lower efficiency.

\subsection{Crosstalk}\label{sec:ctcorr}

In building the VIMOS IFU, the main drivers were the size of the field of 
view and a high sky sampling. This resulted in a large number of spectra 
(400) lying along the 2048 CCD pixels. In this situation, where each fiber 
projects on five pixels on the CCD, neighbouring spectra can contaminate each
other, a phenomenon called ``crosstalk''

The crosstalk  correction must be based on the ``a priori'' knowledge of
the fiber spatial profiles, i.e. the fiber analytic profile and its relevant
parameters must be known from calibration measurements done in the laboratory
on the fiber modules.
For the VIMOS IFU, the fiber profiles
are best described by the combination of three gaussian functions, the first 
one modeling the core of the fiber and the other two, symmetrical with respect
to the central one, modeling the wings.
The positions and widths of the gaussians can then be derived for each 
fiber by fitting the fiber profile on the data themselves, to get the best 
representation of the actual shapes and properly correct for crosstalk.

However, when applying a correction to data, it is unavoidable to introduce an
error. For the correction to be effective, this error must be noticeably 
smaller than the effect going to be corrected. In the case of crosstalk this 
can only be achieved by having very good fits of the fiber profiles in the
cross--dispersion direction, and we note that we have to fit a 3 parameter 
function on a 5 pixels profile.

\begin{figure}
\plotone{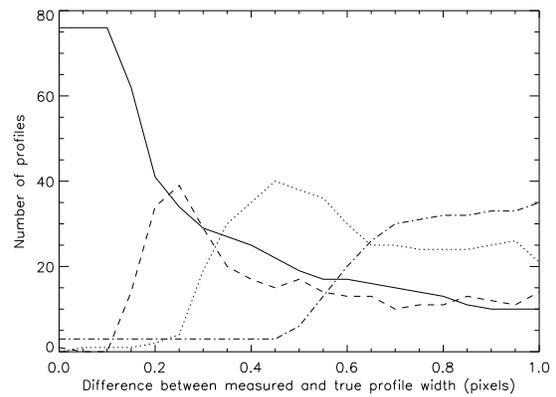}
\caption{Performance of the crosstalk correction as a function of the accuracy
in the measure of the fiber profile width. Solid line: number of fibers with 
error on the recovered flux less than $5\%$. Dashed line: error between $5\%$ 
and $10\%$. Dotted line: error between $10\%$ and $20\%$.
Dot-dashed: error greater than $20\%$.}
\label{fig5}
\end{figure}

To test what is the maximum discrepancy between true and measured values of the
shape parameters which still allows for a good correction of crosstalk, 
simulations have been performed.
It has been found that fiber position uncertainties of about $0.5$ pixels still
guarantee that about $3/4$  of the fibers have an error in measured flux less 
than $5\%$, while the most critical parameter is the profile width, that must 
be known with noticeably higher accuracy. The quality of crosstalk correction
starts to worsen quickly when the maximum error on width measurement becomes of
the order of $0.2$ pixels: in such a case, the number of fibers with an 
accuracy in crosstalk correction worse than $5\%$ rapidly becomes greater than
$50\%$ (see Fig. \ref{fig5}).
Thus, at least for what concerns the profile width, these limiting 
uncertainties on the values of the profile parameters are of the same order of
the accuracy we can achieve in measuring these parameters from real data.

	\placefigure{fig5}

We then estimated what is the level of crosstalk present in the real data. It 
turned out that on average $\sim 5\%$ of the flux of a fiber ``contaminates'' 
each neighbouring fiber, i.e. by applying the correction procedure the 
uncertainties we would introduce in IFU data would be of the same order of 
magnitude as the effect to be corrected.
For this reason the crosstalk correction procedure is not currently implemented
in the IFU data reduction.

\section{Relative Transmission Correction}\label{sec:reltcorr}

Once 1D spectra are traced and extracted from an IFU image, the next
critical step is the correction for fiber relative transmission.
Different fibers have different transmission efficiencies and this effect
alters the intensity of spectra. This is entirely like the effect introduced 
by the pixel to pixel sensitivity variations in direct imaging observations, 
that require flatfielding of the data.

The relative transmission calibration implemented in the VIMOS IFU 
pipeline consists of two steps, both executable at user's choice. First step
is the correction using ``standard'' relative transmission coefficients 
that are provided by ESO as part of the instrument model calibrations
and are usually determined from flat-field calibration frames.
Such a standard correction cannot account for variations in relative 
transmission that may be due for instance to the fact that transmission 
degrades in time, or to variations in instrument position (e.g. rotator 
angle). For this reason a second step calibration is performed: a ``fine'' 
relative transmission correction is executed on the scientific data 
themselves, based on the sky line flux determination described below.
The two steps can be executed in sequence, the second one becoming a refinement
of relative transmission on the data themselves, or just one of them can be
applied to the data. 
This choice guarantees to have the higher degree of flexibility in the data 
reduction procedure, thus allowing the reducer to find the best solution for 
the data under consideration.

The correction for fiber relative transmission is derived by imposing that 
the flux of sky lines must be constant in ALL fibers within one observation,
and that there are no spatial variations inside the IFU field of view.
The flux of a user-selected sky line is computed for each 1D spectrum,
a ``relative transmission'' normalization factor is determined with respect 
to a reference fiber and it is finally applied to the 1D spectra.
Sky line flux may be determined by subtracting from the line intensity the 
contribution of the continuum: regions on both sides of the sky line are 
selected and a second-order polynomial fit is done to obtain the best 
approximation of the continuum underlying the sky line.
As will be noted in Sect. \ref{sec:skysub}, the calibration of the fiber 
relative transmissions with the highest accuracy is of fundamental importance
in order to obtain a good sky subtraction.

\begin{figure*}
\resizebox{\hsize}{!}{\includegraphics{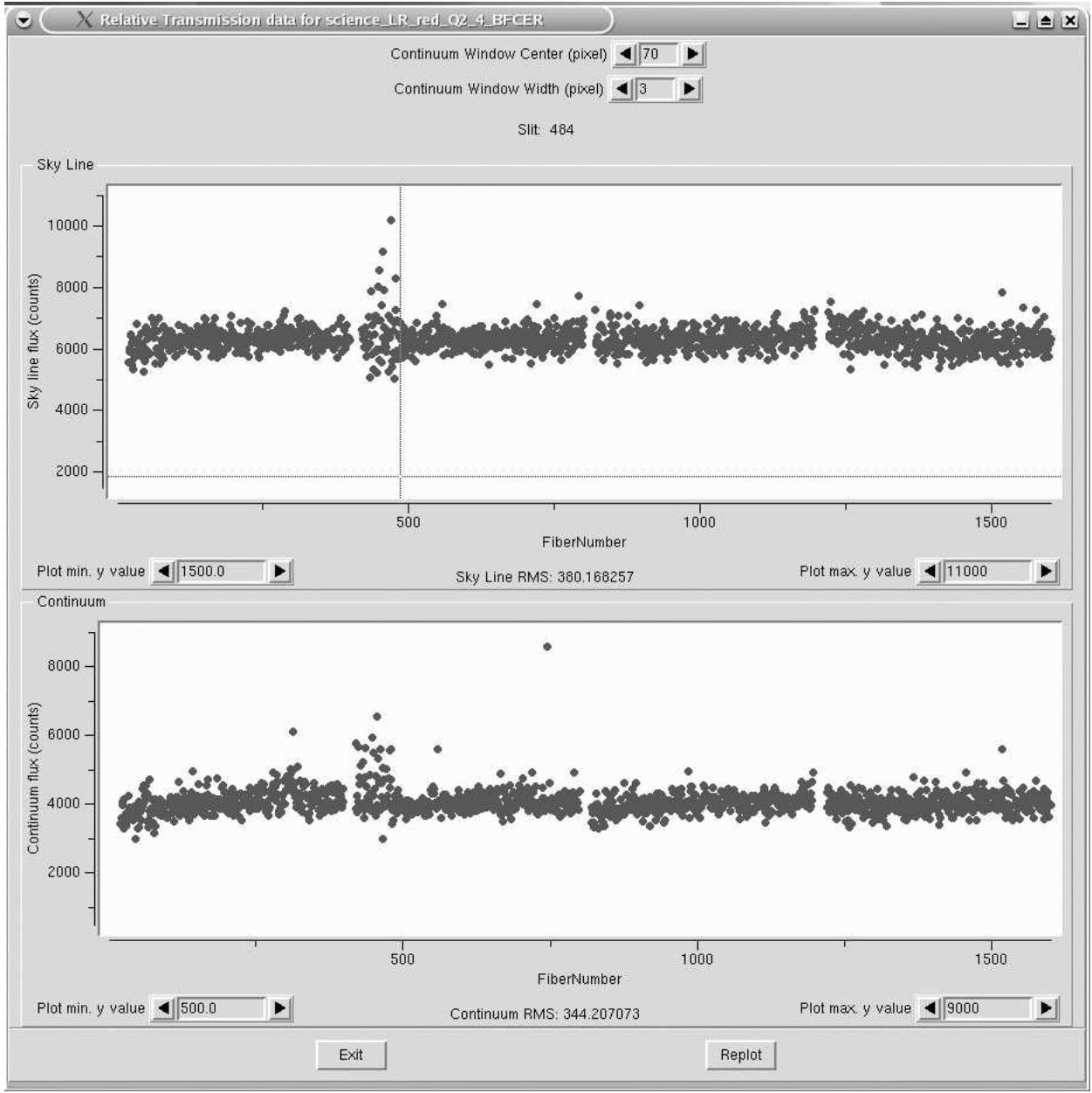}}
\caption{Graphical task for the analysis of the performance of 
relative transmission calibration for IFU data. The top panel shows the 
intensity of the $1600$ spectra at the wavelength of the skyline that has been
used for the calibration (in this example the $5892$ \AA~ line, spectra taken 
with Low Resolution Red grism). 
The bottom panel shows again the spectral intensity, this time computed in a 
user--selected continuum region at $\sim 6000$ \AA. The rms of the intensities
is also shown for each panel.}
\label{fig6}
\end{figure*}

To minimize correction uncertainties, the sky line chosen for calibrating
fine relative transmission coefficients should be characterized by a fairly 
stable flux and be far from the fringing affected regions. In the red domain,
the $5892$ \AA~sky line is usually a good choice, see Fig. \ref{fig6} for an
example. For data taken with the Low Resolution Blue grism, the 
$5577$ \AA~sky line is a very good option.

The presence of uncleaned cosmic rays altering sky line intensities would
lead to an overestimate of the fiber relative transmission, preventing from 
obtaining a correct calibration.
The occurrence of a cosmic ray hit on a given sky line is however extremely 
rare: for observations 45 minutes long, we statistically foresee that only 
$\sim 6$ spectra out of 1600 could be affected by this problem. 
Even if for none of them the cosmic ray removal algorithm cleaned the hits 
on sky lines, it would still be a negligible fraction of spectra.

	\placefigure{fig6}

An interactive plotting task may be used to verify the results of relative 
transmission calibration by looking for residual trends in the data at a 
user-selected wavelength range in the spectra.
An example of the use of this task is shown in Fig. \ref{fig6}, where the 
results of applying the relative transmission calibration on an IFU exposure 
using the $5892$ \AA~ sky line are shown. In the top panel the intensity 
variations of the $5892$ \AA~ line in all the $1600$ spectra of the image are 
plotted, 
while the bottom graph shows the intensity of the continuum at $\sim 6000$ \AA.
Fibers 400 to 480 in Fig. \ref{fig6} belong to a module characterized by 
a very low transmission efficiency, due to a non-optimal assembly of the 
optical components. The low signal from these fibers translates in a noisier
determination of their relative transmission coefficients, and thus in the 
scatter visible in their corrected intensities.

The relative transmission calibration procedure currently implemented in 
the VIPGI pipeline guarantees excellent results, with differences in the
corrected intensities within $5 - 7 \%$ over all modules.

\section{Sky Subtraction}\label{sec:skysub}

Due to its design, the VIMOS IFU does not have ``sky dedicated'' fibers, that 
is fibers that are {\it a priori} known to fall on the sky.
Moreover, once the light enters an optical fiber the information on its spatial
distribution is lost: individual fiber spectra on the CCD do not show regions 
with pure sky emission and regions with object signal, as it happens in slit 
spectroscopy.
Therefore the determination of the sky spectrum to be subtracted from the data,
which is an optional operation, cannot be done in the same way as for classical
spectroscopic data reduction.

IFU spectra can be either the superposition of sky background and 
astronomical object contribution, or pure sky background. In those cases 
where the field of view is relatively empty, i.e. when at least half of the 
fibers do not fall on an object, sky spectrum determination can be achieved 
by properly selecting and combining spectra which are likely to contain pure
sky signal.
These spectra can be selected using the histogram of the total intensities: 
each spectrum is integrated along the wavelength direction, and the total 
flux distribution is built. In a ``mostly empty field'' such distribution 
will show a peak due to pure sky spectra, plus a tail at higher intensities 
where object+sky spectra show up.
Spectra whose integrated intensity is inside a user selected range around the
peak will be pure sky spectra. A median combination of them will ensure that
any residual contamination by faint objects is washed out. 

The sky subtraction method implemented in the  VIPGI pipeline for IFU data
gives good results on deep survey observations of fields devoid of extended
objects, but of course is less optimal for observations of large galaxies
covering the entire field of view.

Physical characteristics of the IFU fibers/lenslets system together with the 
resampling executed on 2D spectra for wavelength calibration cause the fact 
that different fiber spectra are described by different profile shape 
parameters, like FWHM and skewness. 
Along the wavelength direction this reflects in different shapes of the 
spectral lines. Such effect, if not properly taken into account, affects the 
quality of sky subtraction: combining spectra with different line
profiles would lead to an ``average'' sky spectrum whose lines profiles do not
match with any of the original spectra, and subtracting this sky spectrum from
the data would result in the presence of strong s-shape residuals.

\begin{figure}
\plotone{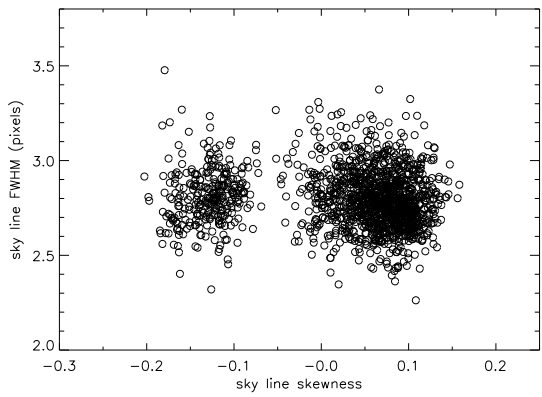}
\caption{Distribution of the $5892$ \AA~ sky line skewness as a
function of sky line width, measured for $\sim 1600$ 1D spectra.}
\label{fig7}
\end{figure}

	\placefigure{fig7}

We performed many tests on real data to determine which are the relevant 
profile parameters: grouping spectra according to the line FWHM value
does not seem to have any influence on the goodness of sky subtraction
(i.e. on the strength of the s-shape residuals).
In Fig. \ref{fig7} the skewness of the $5892$ \AA~ sky line is plotted as a 
function of the line width for $\sim 1600$ spectra. It can be seen from  
Fig. \ref{fig7} that the distribution of line widths is pretty normal, and 
its relatively small dispersion makes it a ``unimportant'' parameter in the 
line profile description. 
Nevertheless, Fig. \ref{fig7} shows instead a very clear bimodality in the 
distribution of line skewness. Such bimodality is surely an artifact deriving
from the undersampling of the line profile: with a typical FWHM of $\sim 3-4$
pixels, we can only see whether skewness is positive or negative. Anyway, our
tests have shown that grouping fibers according to the sign of their skewness
does indeed influence the strength of residuals.
A further classification according to other line shape parameters would lead 
to too many fiber groups, each with too few spectra to allow for a robust
sky determination.
Last, but not least, line shape changes across the focal plane, due to the 
dependence of the instrument focus and optical aberrations on the distance 
from the instrument (in the VIMOS case, quadrant) optical axis. For this 
reason, we analyze separately spectra belonging to different pseudo-slits. 
Thus sky subtraction consists in: 
grouping spectra according to the skewness of a user-selected skyline; for 
spectra belonging to each skewness group the distribution of total 
intensities is built; finally an ``average''  sky spectrum is computed
and subtracted from all the spectra belonging to that group.

\begin{figure*}
\resizebox{\hsize}{!}{\includegraphics{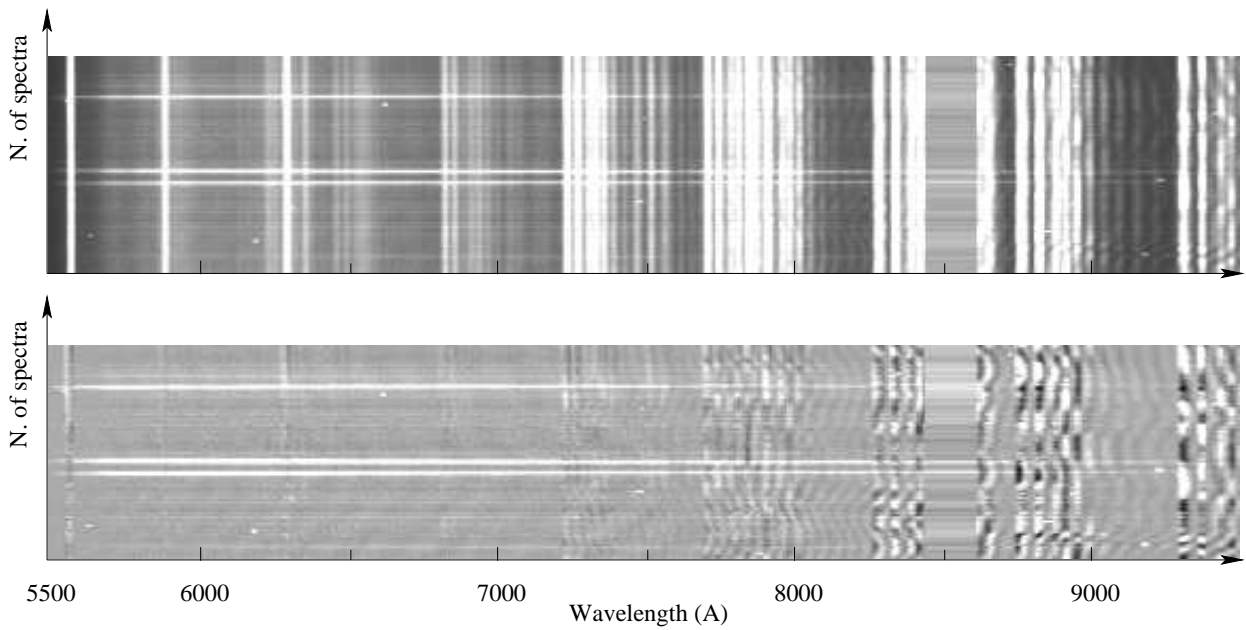}}
\caption{Example of the results obtained by the sky 
subtraction procedure. Top: 1D relative transmission 
corrected spectra in an exposure taken with the Low Resolution Red grism.
Bottom: same spectra after sky subtraction. The $5892$ \AA~ sky line has been
used to group spectra before ``average'' sky spectrum computation. In the
$\sim 8500$ \AA~ region the zero order due  to a nearby
pseudo-slit has been clipped, originating a ``flat'' intensity distribution.}
\label{fig8}
\end{figure*}

Given the adopted method for sky subtraction, an overall good relative 
transmission calibration is essential to get a correct intensity distribution
and thus a proper selection of pure sky spectra.
On the other hand, due to the median combination, if just some spectrum has 
not been perfectly corrected for fiber relative transmission, it will not
affect the sky subtraction step.
This sky subtraction procedure guarantees good results, with the mean level
over the continuum well centered around 0 in spectra where no object signal is
present, and an rms of the order of a few percent. An example of sky subtracted
spectra can be seen in Fig. \ref{fig8}.

	\placefigure{fig8}

\section{Flux calibration}\label{sec:fluxcal}

The last step of data reduction on single observations is flux calibration, 
done in a standard way by multiplying 1D spectra by the sensitivity function 
derived from standard star observations. All the spectra containing flux from 
the standard star are summed together, and the instrument sensitivity function
computed by comparison with the real standard star spectrum from the 
literature.

On the basis of a few objects with known magnitudes that have been observed
with the VIMOS IFU, we have estimated that the overall absolute flux accuracy 
that can be reached with a good quality spectro-photometric calibration 
is of the order of $15\%$, given the unavoidable sources of uncertainty,
such as cross talk contribution (of the order of 5\%), relative transmission 
correction (between 5\% and 10\%) and sky subtraction (few percents).

\section{Final steps: jitter sequences data reduction}\label{sec:finalsteps}

The final result of the single frame reduction procedure is a FITS image,
containing intermediate products of the reduction 
(e.g. spectra not corrected for fiber transmission, or not sky 
subtracted)
under the form of image extensions, plus some tables used and/or created 
during the execution of the different tasks.

In many cases, observations had been carried out using a jittering
technique (the telescope is slightly offsetted from one exposure to the 
next). In these cases, the spectra from the single exposures must be stacked
together according to their offsets,  and in this process also correction for
fringing can be applied.

\subsection{Correcting for fringing}\label{sec:fringcorr}

Due to the characteristics of the VIMOS CCDs, when observing in the red 
wavelength domain, one has to deal with the fringing phenomenon, whose effects
show up at wavelengths larger than $\sim 8200$ \AA~ in VIMOS spectra.

The decision to carry out the fringing correction at the stacking of the 
single exposures stage is dictated by the consideration that during a typical
jitter sequence (a few hours exposure time) the fringing pattern remains 
relatively constant. On the contrary, the overall background intensity and 
the relative strength of the individual sky emission lines can vary 
significantly over the same time scales. Likewise, the physical location of 
the spectra on the CCD changes because of flexures (see Sect. 
\ref{sec:locate} for an evaluation of flexures-induced image motions).
Our approach has been to correct first for the most rapidly changing effects
(first spectra location, secondly sky background) and only in the end to try
to correct for fringing, which is then computed and applied on 1D extracted 
spectra that have already been corrected for cosmic ray hits, for relative 
fiber transmission, etc.

The fringing correction can only be applied on jittered observations and is 
performed separately for the sets of images coming from different quadrants.

First, all the spectra obtained in the jitter sequence for a given fiber are 
combined without taking into account telescope offsets: any object signal is 
thus averaged out in the combination and what remains is a good representation
of the fringing pattern, which is then subtracted from each single spectrum.

\begin{figure*}
\resizebox{\hsize}{!}{\includegraphics{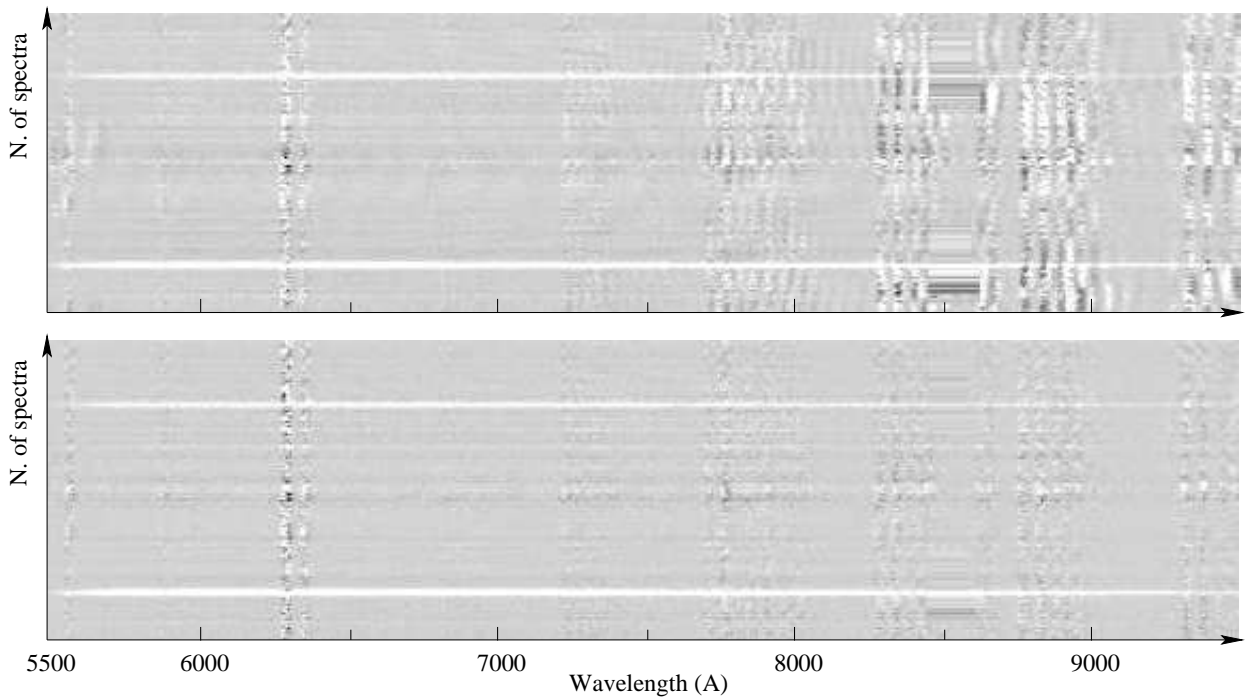}}
\caption{Efficiency of fringing correction: example of 
jitter--combined fully reduced sequence of 9 exposures, 26 minutes each, taken
with the Low Resolution Red grism (wavelength range $5500 - 9500$ \AA) on the 
Chandra Deep Field South. Each row is a different spectrum. Upper and lower 
panels are respectively without and with fringing correction applied to the 
data. As it can be seen in the lower panel, at wavelengths larger than 
$\sim 8200$ \AA~ very low fringing residuals are left after 
correction.}
\label{fig9}
\end{figure*}

The quality of fringing correction is very good: Fig. \ref{fig9} shows the 
result of the reduction of a sequence of 9 jittered exposures on the Chandra 
Deep Field South.
Observations have been done using the Low Resolution Red grism (wavelength 
range $5500 - 9500$ \AA, dispersion $7.14$ \AA /pixel), with single frame
exposure times of $26$ minutes. In  Fig. \ref{fig9}, top, the exposures have 
been combined without applying any correction for fringing, while the bottom 
frame shows the result after having corrected for fringing as explained above. 
It can be seen that fringing correction is efficient in removing almost
completely the residuals in the $\lambda >8200$ \AA~ region of the spectra.

	\placefigure{fig9}

\subsection{Stacking jittered sequences of exposures}\label{sec:jitter}

Stacking is done using all the available images from all the VIMOS quadrants
at a time. In fact, due to the contiguous field of view of the VIMOS IFU and
given how fibers are rearranged on the four VIMOS quadrants, an object 
spectrum can ``move'' from one quadrant to the other going from one jittered
exposure to the next.

Image stacking makes use of datacubes, i.e. 3D images where the (x,y) axes 
sample the spatial coordinates and the z axis samples the wavelength. One 
datacube is created starting from the four images of each IFU exposure.
Jitter offsets are computed by using the header information on telescope 
pointing coordinates or by means of a user-given offset list, and are used
to build a final 3D image starting from single datacubes.
Variations of the relative transmission from quadrant to quadrant, and inside 
the same quadrant from one exposure to the next one in the sequence, are taken
into account by properly rescaling image intensities.

The output of the reduction procedure is a FITS image containing all the 
stacked 1D spectra in the final datacube: each spectrum is written in a row 
of the output image (see Fig. \ref{fig9}), and a correspondence table 
between the position of the spectrum in the final 3D cube appended to it.

\begin{figure*}
\resizebox{\hsize}{!}{\includegraphics{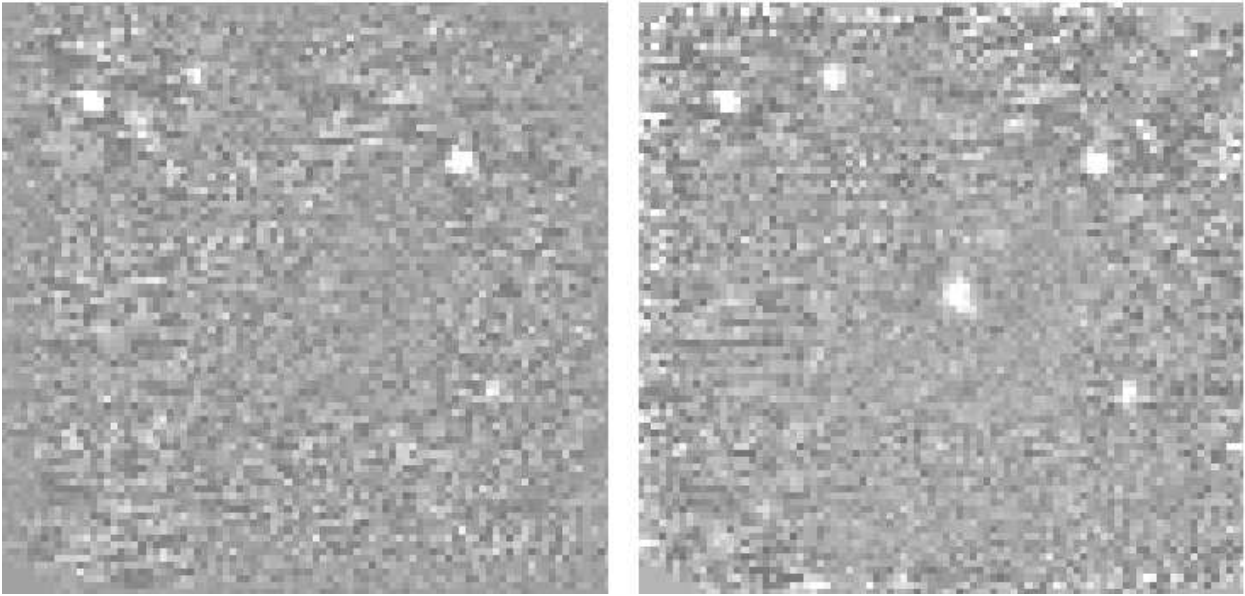}}
\caption{2D reconstructed image of the cluster MRC1022-299 
obtained by integrating over the $5800 \div 8000$ \AA~ wavelength range (left)
and over $100 $ \AA~ centered on the radiogalaxy [OII]3727 emission line 
(right) redshifted at $\simeq 7100$ \AA~ (redshift z=0.9085). The radiogalaxy 
OII emission is clearly extended and asymmetric, and a velocity field can be 
retrieved from the 3D cube. On these images East is up and North is right.
About $3\%$ of the pixels correspond to black/dead fiber spectra and have been
cleaned with the IRAF task {\it fixpix}.}
\label{fig10}
\end{figure*}

	\placefigure{fig10}

Finally, a 2D reconstructed image can be built for scientific analysis.
In Fig. \ref{fig10} we show an example of 2D reconstructed image from VIMOS 
IFU observations of the candidate cluster MRC1022-299 associated with a high
redshift radiogalaxy \citep{mcc96,chap00}.
A sequence of $5$ jittered exposures of $26$ minutes each taken with the Low 
Resolution Red grism has been combined and integrated over the wavelength 
range $5800-8000$ \AA~ (left) and over a narrow band $100$ \AA~ wide 
centered at $7100$ \AA~ (right), where an emission line identified with OII 
is observed in the radiogalaxy spectrum.
The radiogalaxy, invisible in the broad-band image, shows up at the center of
the field in the narrow-band image.
From the OII line we got a redshift of $0.9085$ for the radiogalaxy,
consistent with the value quoted by \citet{mcc96}.

2D reconstructed images can be built from any kind of 1D extracted spectra, 
e.g. transmission corrected or sky subtracted ones, by using the sky-to-CCD 
fiber correspondence given in the standard IFU table or, in the case of a 
jitter sequence reduction, in the associated correspondence table.
These images are particularly useful when dealing with crowded fields data, 
when the automatic sky subtraction recipe cannot guarantee good results. 
In such cases the reduction to 1D spectra can be done without sky 
subtraction.
A preliminary 2D image can be reconstructed and used to interactively 
identify fibers/spectra in object-free regions, to be later combined to 
obtain an accurate estimate of the sky background signal.

\section{Summary}

The VIMOS Integral Field Spectrograph has required a new approach to process 
the large amount of data produced by $6400$ micro-lenses and fibers.

The instrumental IFU setup and the packing of spectra on the $4$ VIMOS 
detectors has  motivated the development of dedicated recipes, with the 
possibility to carefully check the quality of results by means of interactive 
tasks. 

With the large amount of spectra acquired by the VIMOS IFU, it has been 
mandatory to implement the IFU data processing in a pipeline scheme as much 
automated as possible. The VIMOS IFU data processing is implemented under the 
VIPGI environment \citep{Scod05} and is available to the scientific community 
to process VIMOS IFU data since November 2003.

We have estimated that the overall absolute flux accuracy that can be 
reached with our pipeline is of the order of $15\%$, the main sources of 
uncertainty being the cross talk contribution (significant, of the order of 
5\%), the relative transmission correction (between 5\% and 10\%) and sky 
subtraction (few percents).

\acknowledgments

This research has been developed within the framework of the VVDS
consortium.\\
This work has been partially supported by the CNRS-INSU and its Programme 
National de Cosmologie (France), and by Italian Ministry (MIUR) grants
COFIN2000 (MM02037133) and COFIN2003 (num.2003020150).\\
This work has been partly supported by the Euro3D Research Training Network.\\
The VLT-VIMOS observations have been carried out on guaranteed time (GTO) 
allocated by the European Southern Observatory (ESO) to the VIRMOS consortium,
under a contractual agreement between the Centre National de la Recherche 
Scientifique of France, heading a consortium of French and Italian institutes,
and ESO, to design, manufacture and test the VIMOS instrument.

\clearpage

\begin{table*}[t]
\caption{Characteristics of the VIMOS Integral Field Unit.}
\begin{center}
\begin{tabular}{lccccccc}
\multicolumn{1}{l}{Spectral resolution} & \multicolumn{1}{c}{~Field~of~View} 
& Spatial resolution &
Spatial elements & Spectral elements\\ 
    &    &   (arcsec/fiber) &  & (pixel)\\ 
\tableline
Low (R $\sim$ 200)    &  $54^{\prime\prime} \times 54^{\prime\prime}$ & $0.67$ & 6400 & 600 \\
      &        $27^{\prime\prime} \times 27^{\prime\prime}$ &  $0.33$ &    &     \\
High (R $\sim$ 2500)  & $27^{\prime\prime} \times 27^{\prime\prime}$ & $0.67$ & 1600 & 4096 \\
       &        $13^{\prime\prime} \times 13^{\prime\prime}$ & $0.33$ &     &  \\
\end{tabular}
\label{tab:IFU}
\end{center}
\end{table*}

\clearpage
   
\clearpage

\clearpage

%\clearpage

%\clearpage

%\clearpage

\clearpage

%\clearpage

%\clearpage

%\clearpage

%\clearpage

\end{document}